\documentclass[letterpaper]{article}
\usepackage{natbib}
\usepackage{hyperref}
\usepackage[]{graphicx}\usepackage[]{color}
\makeatletter
\def\maxwidth{ %
  \ifdim\Gin@nat@width>\linewidth
    \linewidth
  \else
    \Gin@nat@width
  \fi
}
\makeatother

\definecolor{fgcolor}{rgb}{0.345, 0.345, 0.345}
\newcommand{\hlnum}[1]{\textcolor[rgb]{0.686,0.059,0.569}{#1}}%
\newcommand{\hlstr}[1]{\textcolor[rgb]{0.192,0.494,0.8}{#1}}%
\newcommand{\hlstd}[1]{\textcolor[rgb]{0.345,0.345,0.345}{#1}}%
\newcommand{\hlkwc}[1]{\textcolor[rgb]{0.333,0.667,0.333}{#1}}%
\newcommand{\hlkwd}[1]{\textcolor[rgb]{0.737,0.353,0.396}{\textbf{#1}}}%

\usepackage{framed}
\makeatletter
\newenvironment{kframe}{%
 \def\at@end@of@kframe{}%
 \ifinner\ifhmode%
  \def\at@end@of@kframe{\end{minipage}}%
  \begin{minipage}{\columnwidth}%
 \fi\fi%
 \def\FrameCommand##1{\hskip\@totalleftmargin \hskip-\fboxsep
 \colorbox{shadecolor}{##1}\hskip-\fboxsep
     \hskip-\linewidth \hskip-\@totalleftmargin \hskip\columnwidth}%
 \MakeFramed {\advance\hsize-\width
   \@totalleftmargin\z@ \linewidth\hsize
   \@setminipage}}%
 {\par\unskip\endMakeFramed%
 \at@end@of@kframe}
\makeatother

\definecolor{shadecolor}{rgb}{.97, .97, .97}
\definecolor{messagecolor}{rgb}{0, 0, 0}
\definecolor{warningcolor}{rgb}{1, 0, 1}
\definecolor{errorcolor}{rgb}{1, 0, 0}
\newenvironment{knitrout}{}{} 

\usepackage{alltt}
\usepackage{setspace}

\usepackage{amsmath,amssymb,amsthm}
\usepackage{graphicx}
\usepackage{xspace,soul}
\usepackage{subcaption}

\usepackage{tikz}
\usetikzlibrary{arrows,decorations.pathmorphing,backgrounds,positioning,fit,through,decorations.pathreplacing}

\definecolor{nym-blue}{HTML}{003581}
\definecolor{nym-orange}{HTML}{F47937}
\definecolor{whitesmoke}{HTML}{F5F5F5}

\newcommand{\ds}{MTH 292\xspace}
\newcommand{\sql}{{\sf SQL}\xspace}
\newcommand{\R}{{\sf R}\xspace}

\newcommand{\RM}{{\sf R Markdown}\xspace}
\newcommand{\cmd}[1]{\texttt{#1}}
\newcommand{\bx}{\mathbf{x}}

\newcommand{\Smith}{Smith\xspace}
\newcommand{\Smathies}{Smathies\xspace}

\title{A Data Science Course for Undergraduates:\\Thinking with Data}
\author{Ben Baumer, Smith College}
\IfFileExists{upquote.sty}{\usepackage{upquote}}{}
\begin{document}
\maketitle

\begin{abstract}
Data science is an emerging interdisciplinary field that combines elements of mathematics, statistics, computer science, and knowledge in a particular application domain for the purpose of extracting meaningful information from the increasingly sophisticated array of data available in many settings. These data tend to be non-traditional, in the sense that they are often live, large, complex, and/or messy. A first course in statistics at the undergraduate level typically introduces students with a variety of techniques to analyze small, neat, and clean data sets. However, whether they pursue more formal training in statistics or not, many of these students will end up working with data that is considerably more complex, and will need facility with statistical computing techniques. More importantly, these students require a framework for thinking structurally about data. We describe an undergraduate course in a liberal arts environment that provides students with the tools necessary to apply data science. The course emphasizes modern, practical, and useful skills that cover the full data analysis spectrum, from asking an interesting question to acquiring, managing, manipulating, processing, querying, analyzing, and visualizing data, as well communicating findings in written, graphical, and oral forms. 


\end{abstract}

\section{Introduction}

The last decade has brought considerable attention to the field of statistics, as undergraduate enrollments have swollen across the country. Fueling the interest in statistics is the proliferation of data being generated by scientists, large Internet companies, and seemingly just about everyone. There is widespread acknowledgement---coming naturally from scientists, but also from CEOs and government officials---that these data could be useful for informing decisions. Accordingly, the job market for people who can translate these data into actionable information is very strong, and there is evidence that demand for this type of labor far exceeds supply~\citep{harris2014ds}. By all accounts, students are eager to develop their ability to analyze data, and are wisely investing in these skills. 

But while this data onslaught has strengthened interest in statistics, it has also brought challenges. Modern data streams are importantly different than the data with which many statisticians, and in turn many statistics students, are accustomed to working. For example, the typical data set a student encounters in an introductory statistics course consists of a few dozen rows and three or four columns of non-collinear variables, collected from a simple random sample. These are data that are likely to meet the conditions necessary for statistical inference in a multiple regression model. From a pedagogical point-of-view, this makes both the students and the instructor happy, because the data fits the model, and thus we can proceed to apply the techniques we have learned to draw meaningful conclusions. However, the data that many of our current students will be asked to analyze---especially if they go into government or industry---will not be so neat and tidy. Indeed, these data are not likely to come from an experiment---they are much more likely to be observational. Secondly, they will not likely come in a two-dimensional row-and-column format---they might be stored in a database, or a structured text document (e.g. XML), or come from more than one source with no obvious connecting identifier, or worse, have no structure at all (e.g. data scraped from the web). These data might not exist at a fixed moment in time, but rather be part of a live stream (e.g. Twitter). These data might not even be numerical, but rather consist of text, images, or video. Finally, these data may consist of so many observations, that many traditional inferential techniques might not make sense to use, or even be computationally feasible. 

In 2009, Hal Varian, chief economist at Google, described \emph{statistician} as the ``sexy job in the next 10 years"~\citep{lohr2009stats}. Yet by 2012, the Harvard Business Review used similar logic to declare \emph{data scientist} as the ``sexiest job of the 21st century"~\citep{davenport2012hbr}. Speaking at the 2013 Joint Statistical Meetings, Nate Silver---as always---helped us to unravel what had happened. He noted that ``data scientist is just a sexed up term for a statistician." If Silver is right, then the statistics curriculum should be updated to include topics that are currently more closely associated with data science than with statistics (e.g. data visualization, database querying, algorithmic concerns about computation techniques)\footnote{The Wikipedia defines ``data science" as ``the extraction of knowledge from data", whereas ``statistics" is "the study of the collection, analysis, interpretation, presentation, and organization of data." Does writing an SQL query belong to both?}. It is clear that statisticians and data scientists, broadly, share a common goal---namely, to use data appropriately to inform decision-making---but what we describe in this paper is a course at a small liberal arts college in \emph{data science} that is atypical within the current statistics curriculum. Nevertheless, what we present here is wholly consistent with the vision for the future of the undergraduate statistics curriculum articulated by~\cite{horton2015challenges}. The purpose of this course is to prepare students to work with these modern data streams as described above. Some of the topics covered in this course have historically been the purview of computer science. But while the course we describe indisputably contains elements of statistics and computer science, it just as indisputably belongs exclusively to neither discipline. Furthermore, it is not simply a collection of topics from existing courses in statistics and computer science, but rather an integrated presentation of something more holistic.




\section{Background and Related Work}

While many believe that to understand statistical theory, a solid foundation in mathematics is necessary, it seems clear that \emph{computing} skills are necessary for one to become a functional, practicing statistician. In making this analogy~\cite{nolan2010computing} argued strongly for a larger presence for computing in the statistics curriculum. Citing this work, the~\cite{asa-guidelines} underscored the importance of computing skills (even using the words ``data science") in the 2014 guidelines for undergraduate majors in statistical science. Here, by \emph{computing}, we mean \emph{statistical programming} in an environment such as \R. It is important to recognize this as a distinct---and more valuable---skill than being able to perform statistical computations in a menu-and-click environment such as Minitab. Indeed, \cite{nolan2010computing} go even further, advocating for the importance of teaching general command-line programs, such as \cmd{grep} (for regular expressions) and other common UNIX commands that really have nothing to do with statistics, \emph{per se}, but are incredibly useful for cleaning and manipulating documents of many types. 

Although practicing statisticians seem to largely agree that the lion's share of the time spent on many projects is devoted to data cleaning and manipulation (or \emph{data wrangling}, as it is often called~\citep{kandel2011research}), the motivation for adding these skills to the statistics curriculum is not simply convenience, nor should a lack of skills or interest on the part of instructors stand in the way. \cite{finzer2013data} describes a ``data habit of mind...that grows out of working with data." (This is not to be confused with ``statistical thinking" as articulated by~\cite{chance2002st}, which contains no mention of computing.)  
In this case, a data habit of mind comes from experience working with data, and is manifest in people who start thinking about data formatting \emph{before} data gets collected~\citep{zhu2013data}, and have a foresight about how data should be stored that is informed by how it will be analyzed. Furthermore, while some might view \emph{data management} as a perfunctory skill on intellectual par with \emph{data entry}, there are others thinking more broadly about data.\footnote{Examples of poor data management abound, but one of the most common is failure to separate the actual data from the analysis of that data. Microsoft Excel is a particular villian in this arena, where merged cells, rounding induced by formatted columns, and recomputed formulas can result in the ultimate disaster: losing the original recorded data!} Just as~\cite{wilkinson2006grammar} brought structure to graphics through ``grammar," \cite{tidy-data,dplyr} brought structure to data through ``verbs." These common data manipulation techniques are the practical descendents of theoretical work on data structures by computer scientists who developed notions of normal forms, relational algebras, and database management systems. 


While the emphasis on computing within the statistics curriculum may be growing, it belongs to a larger, more gradual evolution in statistics education towards data analysis---with computers, and encourages us to reflect on shifting boundaries between statistics and computer science. \cite{moore1998statistics}---viewing statistics as an ongoing quest to ``reason about data, variation, and chance"---saw statistical thinking as a powerful anchor that would prevent statistics from being ``overwhelmed by technology." Cobb has argued both for an increased emphasis on conceptual topics in statistics~\citep{cobb2011ts}, but also sees the development of statistical theory as an anachronistic consequence of a lack of computing power~\citep{cobb:2007}. Moreover, while much of statistical theory was designed to make the strongest inference possible from what was often scarce data, our current challenge is trying to extract anything meaningful from abundant data. \cite{breiman2001statistical} articulated the distinction between ``statistical data models" and ``algorithmic models" that in many ways characterizes the relationship between statistics and machine learning, viewing the former as being far more limited than the latter. And while machine learning and data mining have traditionally been subfields of computer science, \cite{finzer2013data} notes that data science does not have a natural home within traditional departments, belonging exclusively to neither mathematics, statistics, or computer science. Indeed, in~\cite{cleveland2001data}'s seminal action plan for data science, he saw data science as a ``partnership" between statisticians (i.e. data analysts) and computer scientists. 




\section{The Course}

In this paper we describe an experimental course taught at \Smith College in the fall of 2013 and again in the fall of 2014 called MTH 292: Data Science. In the first year, 18 students completed the course, as did another 24 in the following year. The prerequisites were an introductory statistics course and some programming experience. Existing courses at the University of California-Berkeley, as well as Macalester and St. Olaf Colleges, are the pedagogical cousins of \ds. 

\ds can be separated into a series of two-to-three week modules: data visualization, data manipulation/data wrangling, computational statistics, machine learning (or statistical learning), additional topics. In what follows we provide greater detail on each of these modules.

\paragraph{Learning Outcomes}

\begin{figure}
  \centering
    		\begin{tikzpicture}[xscale=1.5,yscale=1.5
				, arrow/.style={->,draw=nym-orange,line width=1pt,bend angle=90}
  			, stats-arrow/.style={->,draw=nym-blue,line width=1pt,bend angle=45,dashed}
				, endpoint/.style={draw=nym-blue,very thick,text width=2.8cm,text centered,fill=whitesmoke,text=nym-blue,font=\fontsize{7}{9}\selectfont}
				, popup/.style={rounded corners,draw=nym-blue,very thick,text width=2.8cm,text centered,fill=whitesmoke,text=nym-blue,font=\fontsize{7}{9}\selectfont}
				, ds/.style={rounded corners,draw=nym-orange,very thick,text width=2.8cm,text centered,fill=whitesmoke,text=nym-blue,font=\fontsize{7}{9}\selectfont}
				]
				\node[endpoint] (question) at (0,6) {Question};
				\node[popup] (data) at (0,5) {Data};
				\node[popup] (stat) at (0,4) {Methods};
				\node[popup] (interpret) at (0,3) {Inference};
				\node[popup] (present) at (0,2) {Presentation};
				\node[endpoint] (answer) at (0,1) {Answer};
				\node[ds] (acq) at (3,6) {Data Acquisition};
				\node[ds] (clean) at (3,5.5) {Data Processing\\Data Cleaning};
				\node[ds] (manage) at (3,5) {Data Management};
				\node[ds] (store) at (3,4.5) {Data Storage};
				\node[ds] (retrieve) at (3,4) {Data Retrival};
				\node[ds] (mine) at (-3,4.25) {Data Mining\\Machine Learning};
				\node[ds] (compute) at (-3,3) {Computational Statistics};
				\node[ds] (compute2) at (-3,3.75) {Regression};
				\node[ds] (viz) at (3,3) {Visualization};
				\node[ds] (graphic) at (3,2.5) {Data Graphic Design};
				\node[ds] (oral) at (3,2) {Oration};
  			\path[stats-arrow] (question) edge (data);
  			\path[stats-arrow] (data) edge (stat);
				\path[stats-arrow] (stat) edge (interpret);
				\path[stats-arrow] (interpret) edge (present);
				\path[stats-arrow] (present) edge (answer);
				\path[arrow] (question) edge (acq);
				\path[arrow] (acq) edge (clean);
				\path[arrow] (clean) edge (manage);
				\path[arrow] (manage) edge (store);
				\path[arrow] (store) edge (retrieve);
				\path[arrow] (retrieve) edge (data);
				\path[arrow] (data) edge[out=210,in=150] (stat);
				\path[arrow] (mine) edge (stat);
				\path[arrow] (compute2) edge (stat);
				\path[arrow] (compute) edge (interpret);
  			\path[arrow] (stat) edge[out=210,in=150] (interpret);
				\path[arrow] (interpret) edge (viz);
				\path[arrow] (viz) edge (graphic);
				\path[arrow] (graphic) edge (oral);
				\path[arrow] (oral) edge (present);
				\path[arrow] (present) edge[out=210,in=150] (answer);
			\end{tikzpicture}
	\caption{Schematic of the modern statistical analysis process. The introductory statistics course (and in many cases, the undergraduate statistics curriculum) emphasizes the central column. In this data science course, we provide instruction into the bubbles to the left and right.}
	\label{fig:data-science}
\end{figure}
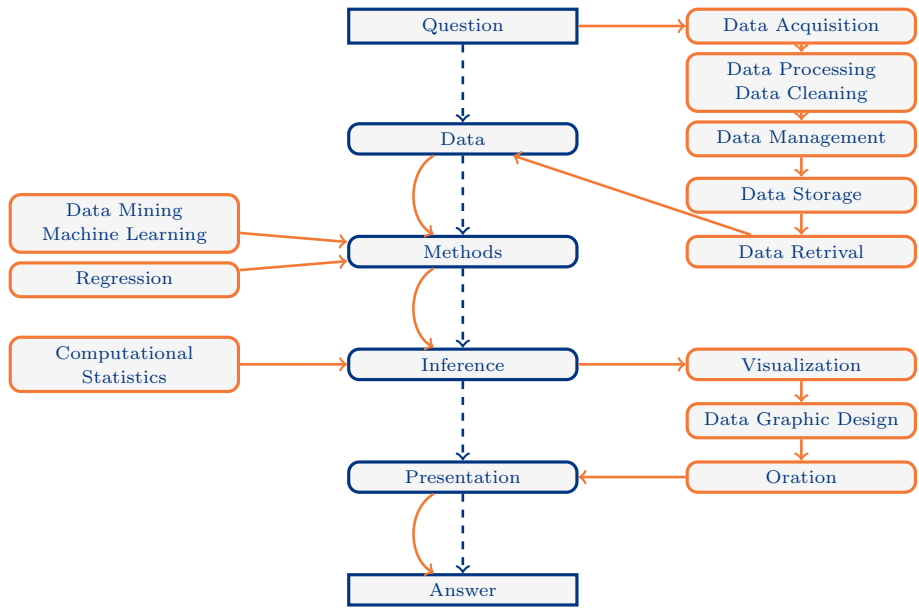

In Figure \ref{fig:data-science}, we present a schematic of a modern statistical analysis process, from forming a question to obtaining an answer. In the introductory statistics course, we teach a streamlined version of this process, wherein challenges with the data, computational methods, and visualization and presentation are typically not taught. These processes inform the material presented in the data science course. The goal is to produce students who have \emph{confidence} and foundational skills---not necessarily expertise---to tackle each step in this modern data analysis cycle, both immediately and in their future careers. 

\subsection{Day One}

The first class provides an important opportunity to hook students into data science. Since most students do not have a firm grasp of what data science is, and in particular, how it differs from statistics, Figure \ref{fig:data-science} can help draw these distinctions. The goal is to illustrate the richness and vibrance of data science, and emphasize its inclusiveness by highlighting the different skills necessary for each task. Students should be sure within the first five minutes of the semester that there is something interesting and useful for them to learn in the course. 

Next, we engage students immediately by exposing them to a recent, relevant example of data science. In the fall of 2013, we chose a provocative paper by~\cite{digrazia2013} that was under review at the time. Additionally, students are asked to read a rather ambitious editorial in \textit{The Washington Post} written by one of the authors, a sociologist, in which he claims that Twitter will put political pollsters out of work~\citep{rojas2013}. 

This is a typical data science research project, in that:
\begin{itemize}
  \itemsep0in
  \item The data being analyzed were scraped from the Internet, not collected from a survey or clinical trial. Typical statistical assumptions about random sampling are clearly not met. 
  \item The research question was addressed by combining \emph{domain knowledge} (i.e. knowledge of how Congressional races work) with a data source (Twitter) that had no obvious relevance to one another.
  \item A \emph{large} amount of data (500 million tweets!) was collected (although only 500,000 tweets were analyzed)---so large that the data itself was a challenge to manage. In this case, the data were big enough that the help of the Center for Complex Networks and Systems Research at Indiana University was enlisted.
  \item The project was undertaken by a team of researchers from different fields (i.e. sociology, computing) working in different departments, and bringing different skills to bear on the problem---a paradigm that many consider to be optimal.
\end{itemize}

Students are then asked to pair up and critically review the paper. The major findings reported by the authors stem from the interpretation of two scatterplots and two multiple regression models, both of which are accessible to students who have had an introductory statistics course. There are several potential weaknesses in both the plots presented in the paper~\citep{linkins2013, gelman2013}, and the interpretation of the coefficients in the multiple regression model, which some students will identify. The exercise serves to refresh students' memories about statistical thinking, encourage them to think critically about the display of data, and illustrate the potential hazards of drawing conclusions from data in the absence of a statistician. Instructors could also use this discussion as a segue to topics in experimental design, or introduce the ASA's Ethical Guidelines for Statistical Practice~\citep{asa-ethics}.

Finally, students are asked quite literally about how they would go about reproducing this study. That is, they are asked to identify all of the steps necessary to conduct this study, from collecting the data to writing the paper, and to think about whether they could accomplish this with their current skills and knowledge. While students are able to generate many of the steps as a broad outline, most are unfamiliar with the practical considerations necessary. For example, students recognize that the data must be downloaded from Twitter, but few have any idea how to do that. This leads to the notion of an API (application programming interface), which is provided by Twitter (and can be used in several environments, notably \R and Python). Moreover, most students do not recognize the potential difficulties of storing 500 million tweets. How big is a tweet? Where and how could you store them? Spatial concerns also arise: how do you know in which Congressional district the person who tweeted was? Most students in the class have experience with \R, and thus are comfortable building a regression model and overlaying it on a scatterplot. But few have considered anything beyond the default plotting options. How do you add annotations to the plot to make it more understandable? What are the principles of data graphic design that would lead you to think certain annotations are necessary or appropriate? 

Students are then advised that this course will give them the tools necessary to carry out a similar study. This will involve improving their skills with programming, data management, data visualization, and statistical computing. The goal is to leave students feeling \emph{energized}, but open to exploring their newly-acquired, more complex understanding of data.

\subsection{Data Visualization}

From the first day of class, students are reminded that statistical work is of limited value unless it can be communicated to non-statisticians~\citep{swires2014pd}. More specifically, most data scientists working in government or industry (as opposed to those working in academia) will work for a boss who possesses less technical knowledge than she. A perfect, but complicated, statistical model may not be persuasive to non-statisticians if it cannot be communicated clearly. Data graphics provide a mechanism for illustrating relationships among data, but most students have never been exposed to structured ideas about how to create effective data graphics. 

In \ds, the first two weeks of class are devoted to data visualization. This serves two purposes: 1) it is an engaging hook for a science course in a liberal arts school; and 2) it gives students with weaker programming backgrounds a chance to get comfortable in \R. 

Students read the classic text of \cite{tufte1983visual} in its entirety, as well as excerpts from \cite{yau2013dp}. The former provides a wonderfully cantankerous account of what \emph{not} to do when creating data graphics, as well as thoughtful analyses of how data graphics should be constructed. My students took delight in critiquing data graphics that they had found online through the lens crafted by Tufte. The latter text, along with \cite{yau2011visualize}, provides many examples of interesting data visualizations that can be used in the beginning of class to inspire students to think broadly about what can be done with data (e.g. \emph{data art}). Moreover, it provides a well-structured taxonomy for composing data graphics that give students an orientation into data graphic design. For example, in Yau's taxonomy, a data graphic that uses color as a visual cue in a Cartesian coordinate system is what we commonly call a heat map. Students are also exposed to the hierarchy of visual perception that stems from work by \cite{cleveland2001data}. 

Homework questions from this part of the course focus on demonstrating understanding by critiquing data graphics found ``in the wild," an exercise that builds \emph{confidence} (i.e. ``Geez, I already know more about data visualization that this guy..."). Computational assignments introduce students to some of the more non-trivial aspects of annotating data graphics in \R (e.g. adding textual annotations and manipulating colors, scales, legends, etc.). We discuss additional topics in data visualization in Section \ref{sec:topics}. 

\subsection{Data Manipulation/Data Wrangling}

As noted earlier, it is a common refrain among statisticians that ``cleaning and manipulating the data" comprises an overwhelming majority of the time spent on a statistical project. In the introductory class, we do everything we can to shield students from this reality, exposing them only to carefully curated data sets. By contrast, in \ds students are expected to master a variety of common data manipulation techniques. The term \emph{data management} has a boring, IT connotation, but there is a growing acknowledgement that such \emph{data wrangling}, or \emph{data manipulation} skills are not only valuable, but in fact belong to a broader intellectual discipline~\citep{tidy-data}. One of the primary goals of \ds is to develop students' capacity to ``think with data"~\citep{nolan2010computing}, in both a practical and theoretical sense. 

Over the next three weeks, students are given rapid instruction in data manipulation in \R and \sql. In the spirit of the data manipulation ``verbs" advocated by~\cite{dplyr}, students learn how to perform the most fundamental data operations in both \R and \sql, and are asked to think about their connection. 

\begin{itemize}
  \itemsep0in
  \item \emph{select}: subset variables (\cmd{SELECT} in SQL, \cmd{select()} in \R (\cmd{dply}))
  \item \emph{filter}: subset rows (\cmd{WHERE, HAVING} in SQL, \cmd{filter()} in \R)
  \item \emph{mutate}: add new columns (\cmd{... AS ...} in SQL, \cmd{mutate()} in \R)
  \item \emph{summarise}: reduce to a single row (\cmd{GROUP BY} in SQL, \cmd{summarise(group\_by())} in \R)
  \item \emph{arrange}: re-order the rows (\cmd{ORDER BY} in SQL, \cmd{arrange()} in \R)
\end{itemize}

By the end, students are able to see that an \sql query containing 
  \begin{verbatim}
    SELECT ... FROM a JOIN b WHERE ... GROUP BY ... HAVING ... ORDER BY ...
  \end{verbatim} 
  is equivalent to a chain of \R commands involving 
  \begin{verbatim}
    a %>%
      select(...) %>%
      filter(...) %>%
      inner_join(b, ...) %>%
      group_by(...) %>%
      summarise(...) %>%
      filter(...) %>%
      arrange(...)
  \end{verbatim}

A summary of analogous \R and \sql syntax is shown in Table \ref{tab:sql-r}.

Moreover, students learn to determine for themselves, based on the attributes of the data (most notably size), which tool is more appropriate for the type of analysis they wish to perform. They learn that \R stores data in memory, so that the size of the data with which you wish to work is limited by the amount of memory available to the computer, whereas \sql stores data on disk, and is thus much better suited for storage of large amounts of data. However, students learn to appreciate the virtually limitless array of operations that can be performed on data in \R, whereas the number of useful computational functions in \sql is limited. Thus, students learn to make choices about software in the context of hardware---and data.

\begin{table}
  \centering
  \begin{tabular}{|c|p{5cm}|p{5cm}|}
  \hline
  Concept & SQL & \R (\cmd{dplyr}) \\
  \hline
  Filter by rows \& columns 
    & \cmd{SELECT col1, col2 FROM $a$ WHERE col3 = 'x'} 
    & \cmd{select(filter($a$, col3 == "x"), col1, col2)} \\
  \hline
  Aggregate by rows 
    & \cmd{SELECT id, sum(col1) as total FROM $a$ GROUP BY id} 
    & \cmd{summarise(group\_by($a$, id), total = sum(col1))} \\
  \hline
  Combine two tables 
    & \cmd{SELECT * FROM $a$ JOIN $b$ ON a.id = b.id} 
    & \cmd{inner\_join(x=$a$, y=$b$, by="id"))} \\
  \hline
  \end{tabular}
  \caption{Conceptually analogous SQL and \R commands. Suppose $a$ and $b$ are SQL tables or \R \cmd{data.frame}s}
  \label{tab:sql-r}
\end{table}

%
%
%
%

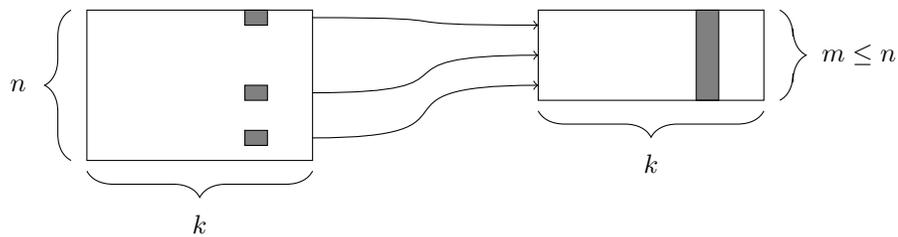
\begin{figure}
  \centering
    		\begin{tikzpicture}[xscale=3,yscale=2,
				coverage/.style={circle,draw=blue!50,fill=blue!20,thick,opacity=0.5},
				lifetime/.style={->,dashed,thick,gray},
				sensor/.style={circle,draw=black,fill=red!50,thick,opacity=0.75,inner sep=0pt,minimum size=2mm},
				nrow/.style={decoration={brace,amplitude=10pt},xshift=-2pt,yshift=0pt},
				ncol/.style={decoration={brace,amplitude=10pt},xshift=0pt,yshift=-2pt}
				]
				\draw[fill=gray] (0.7,0.9) rectangle (0.8,1);
				\draw[fill=gray] (0.7,0.4) rectangle (0.8,0.5);
				\draw[fill=gray] (0.7,0.1) rectangle (0.8,0.2);
				\draw [] (0,0) rectangle (1, 1);
				\draw [decorate,nrow] (0,0) -- (0,1) node [midway,xshift=-20pt] {$n$};
				\draw [decorate,ncol] (1,0) -- (0,0) node [midway,yshift=-20pt] {$k$};

				\draw[fill=gray] (2.7,0.4) rectangle (2.8,1);
				\draw [] (2,0.4) rectangle (3, 1);
				\draw [decorate,nrow,xshift=+4pt] (3,1) -- (3,0.4) node [midway,xshift=30pt] {$m \leq n$};
				\draw [decorate,ncol] (3,0.4) -- (2,0.4) node [midway,yshift=-20pt] {$k$};
				\draw[->] (1, 0.95) to [out = 0, in = 180, looseness = 2] (2,0.9);
				\draw[->] (1, 0.45) to [out = 0, in = 180, looseness = 2] (2,0.7);
				\draw[->] (1, 0.15) to [out = 0, in = 180, looseness = 2] (2,0.5);
			\end{tikzpicture}
      \caption{The subset operation}
      \label{fig:subset}
\end{figure}

\begin{figure}
  \centering
    		\begin{tikzpicture}[xscale=3,yscale=2,
				coverage/.style={circle,draw=blue!50,fill=blue!20,thick,opacity=0.5},
				lifetime/.style={->,dashed,thick,gray},
				sensor/.style={circle,draw=black,fill=red!50,thick,opacity=0.75,inner sep=0pt,minimum size=2mm},
				nrow/.style={decoration={brace,amplitude=10pt},xshift=-2pt,yshift=0pt},
				ncol/.style={decoration={brace,amplitude=10pt},xshift=0pt,yshift=-2pt}
				]
				\draw [] (0,0) rectangle (1, 1);
				\draw [] (1/3,0) rectangle (2/3, 1);
				\coordinate [label=$x$] (x) at (1/6, 0.5);
				\coordinate [label=$y_1$] (y1) at (0.5, 0.5);
				\coordinate [label=$y_2$] (y2) at (5/6, 0.5);
				\coordinate [label=above:$v_1$] (v1) at (0.5, 1);
				\coordinate [label=above:$v_2$] (v2) at (5/6, 1);
				\draw [decorate,nrow] (0,0) -- (0,1) node [midway,xshift=-20pt] {$n$};
				\draw [decorate,ncol] (1,0) -- (0.4,0) node [midway,yshift=-20pt] {$k$};

				\draw[->,line width=4pt] (1.2, 0.5) -- (1.6, 0.5);

				\draw [] (2,0) rectangle (3,1);
				\draw [] (2,-1) rectangle (3, 0);
				\draw [] (2 + 1/3, -1) rectangle (2 + 2/3, 1);
				\coordinate [label=$x$] () at (2 + 1/6, 0.5);
				\coordinate [label=$x$] () at (2 + 1/6, -0.5);
				\coordinate [label=$y_1$] () at (2 + 5/6, 0.5);
				\coordinate [label=$y_2$] () at (2 + 5/6, -0.5);
				\coordinate [label=$v_1$] () at (2.5, 0.5);
				\coordinate [label=$v_2$] () at (2.5, -0.5);
				\draw [decorate,nrow] (2,-1) -- (2,1) node [midway,xshift=-20pt] {$nk$};
				\draw [decorate,ncol] (3,-1) -- (2 + 1/3,-1) node [midway,yshift=-20pt] {$2$};
			\end{tikzpicture}
      \caption{The gather operation}
      \label{fig:reshape}
  \end{figure}
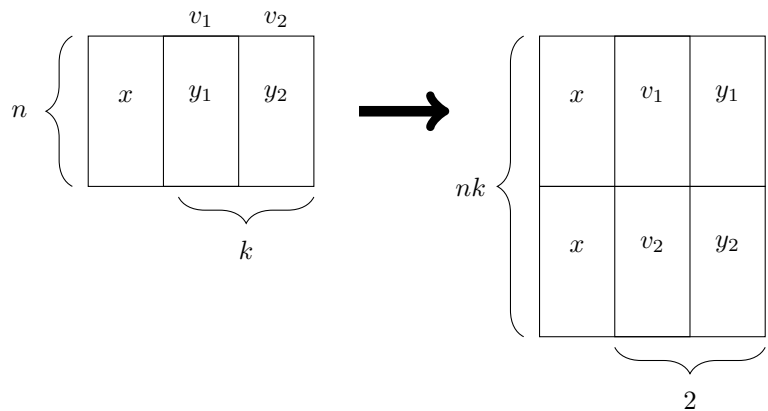

Care must be taken to make sure that what students are learning at this stage of the course is not purely progamming syntax (although that is a desired side effect). Rather, they are learning more generally about operations that can be performed on data, in two languages. To reinforce this, students are asked to think about a physical representation of what these operations do. For example, Figure \ref{fig:subset} illustrates conceptually what happens when row filtering is performed on a \cmd{data.frame} in \R or a table in \sql. Less trivially, Figure \ref{fig:reshape} illustrates the incredibly useful \cmd{gather} operation in \R. 

\subsection{Computational Statistics}

Now having the intellectual and practical tools to work with data and visualize it, the third part of the course provides students with computational statistical methods for analyzing data in the interest of answering a statistical question. There are two major objectives for this section of the course:

\begin{enumerate}
  \item Developing comfort constructing interval estimates using resampling techniques (e.g. the bootstrap). Understanding the nature of variation in observational data and the benefit of presenting interval estimates over point estimates.
  \item Developing comfort with OLS linear regression, beginning with simple linear regression (which all students should have already seen in their intro course), but continuing to include multiple and logistic regression, and a few techniques for automated feature selection. 
\end{enumerate}

The first objective underlines the \emph{statistical} elements of the course, encouraging students to put observations in relevant context by demonstrating an understanding of variation in their data. The second objective, while not a substitute for a semester course in regression analysis, helps reinforce a practical understanding of regression, and sets the stage for the subsequent machine learning portion of the course. 

\subsection{Machine Learning}

Two weeks were devoted to introductory topics in machine learning. Some instructors may find that this portion of the course overlaps too heavily with existing offerings in computer science or applied statistics. Others might argue that these topics will not be of interest to students who are primarily interested in the softer side of data science. However, a brief introduction to machine learning gives students a functional framework for testing algorithmic models. Assignments force them to grapple with the limitations of large data sets, and pursue statistical techniques that are beyond introductory. 

In order to understand machine learning, one must recognize the differences between the mindset of the data miner and the statistician, notably characterized by~\cite{breiman2001statistical}, who distinguished two types of models $f$ for $y$, the response variable, and $\bx$, a vector of explanatory variables. One might consider a \emph{data model} $f$ such that $y \sim f(\bx)$, assess whether $f$ could reasonably have been the process that generated $y$ from $\bx$, and then make inferences about $f$. The goal here is to learn about the real process that generated $y$ from $\bx$, and the conceit is that $f$ is a meaningful reflection of the true unknown process. Alternatively, one might construct an \emph{algorithmic model} $f$, such that $y \sim f(\bx)$, and use $f$ to predict unobserved values of $y$. If it can be determined that $f$ does in fact do a good job of predicting values of $y$, one might not care to learn much about $f$. In the former case, since we want to learn about $f$, a simpler model may be preferred. Conversely, in the latter case, since we want to predict new values of $y$, we may be indifferent to model complexity (other than concerns about overfitting and scalability). 

These are \emph{very} different perspectives to take towards learning from data, so after reinforcing the former perspective that students learned in their introductory course, \ds students are exposed to the latter point-of-view. These ideas are further explored in a class discussion about Chris Anderson's famous article on \textit{The End of Theory}~\citep{anderson2008}, in which he argues that the abundance of data and computing power will eliminate the need for scientific modeling.

The notions of cross-validation and the confusion matrix frame the machine learning unit (ROC curves are also presented as an evaluation technique). The goal is typically to predict the outcome of a binary response variable. Once students understand that these predictions can be evaluated using a confusion matrix, and that models can be tested via cross-validation schemes, the rest of the unit is spent learning classification techniques. The following techniques were presented, mainly at a conceptual and practical level: decision/classification trees, random forests, $k$-nearest neighbor, n\"{a}ive Bayes, aritificial neural networks, and ensemble methods.

One of the most satisfying aspects of this unit is that you can now turn students loose on a massive data set. Past instances of the KDD Cup (\url{http://www.sigkdd.org/kddcup/index.php}) are an excellent source for such data sets. We explored data from the 2008 KDD Cup on breast cancer. Each of the $n$ observations contained digitized data from an X-Ray image of a breast. Each observation corresponded to a small area of a particular breast, which may or may not depict a malignant tumor---this provided the binary response variable. In addition to a handful of well-defined variables ($(x,y)$-location, etc.), each observation has 117 nameless attributes, about which no information was provided. Knowing nothing about what these variables mean, students recognized the need to employ machine learning techniques to sift through them and find relationships. The size of the data and number of variables made manual exploration of the data impractical. 

Students were asked to take part in a multi-stage machine learning ``exam"~\citep{cohen1998pyramid} on this breast cancer data. In the first stage, students had several days to work alone and try to find the best logistic regression model that fit the data. In the second stage, students formed groups of three, discussed the strengths and weaknesses of their respective models, and then built a classifier, using any means available to them, that best fit the data. The third stage of the exam was a traditional in-class exam.  

\subsection{Additional Topics}
\label{sec:topics}

As outlined above, data visualization, data manipulation, computational statistics, and machine learning comprise the four pillars of this data science course. However, additional content can be layered in at the instructor's discretion. We list a few such topics below. Greater detail is provided in our supplementary materials. 

\begin{itemize}
  \itemsep0in
  \item Spatial Analysis: creating appropriate and meaningful graphical displays for data that contains geographic coordinates
  \item Text Mining \& Regular Expressions: learning how to use regular expressions to produce data from large text documents
  \item Data Expo: exposing students to the questions and challenges that people outside the classroom face with their own data
  \item Network Science: developing methods for data that exists in a network setting (i.e. on a graph)
  \item Big Data: illustrating the next frontier for working with data that is truly large scale
\end{itemize}

\section{Computing}

Practical, functional programming ability is essential for a data scientist, and as such no attempt is made to shield students from the burden of writing their own code. Copious examples are given, and detailed lecture notes containing annotated computations in \R are disseminated each class. Lectures jump between illustrating concepts on the blackboard and writing code on the computer projected overhead, and students are expected to bring their laptops to class each day and participate actively. While it is true that many of the students struggled with the programming aspect of the course, even those that did expressed enthusiasm and satisfaction as they became more comfortable. Newly focused on becoming data scientists, several students went on to take subsequent courses on data structures or algorithms offered by the computer science department during the next semester. 

In this course, programming occured exclusively in \R and SQL. Others may assert that Python is necessary, and future incarnations of this course may include more Python. In my view these are the three must-have languages for data science.~\footnote{SQL is a mature technology that is widely-used, but useful for a specific purpose. \R is a flexible, extensible platform that is specifically designed for statistical computing, and represents the current state-or-the-art. Python has become something of a \emph{lingua franca}, capable of performing many of the data analysis operations otherwise done in \R, but also being a full-fledged general purpose programming language with lots of supporting packages and documentation. 

At \Smith, all introductory computer science students learn Python, and all introductory statistics students in the mathematics and statistics department learn \R. However, it is not clear yet how large the intersection of these two groups is. It is probably easier for those who know Python to learn \R than it is for those who know \R to learn Python, and thus the decision was made in this instance to avoid Python and focus on \R. Other instructors may make different choices without disruption.}

\section{Assignments}

Reading assignments in \ds were culled from a variety of textbooks and articles available for free online. Non-trivial sections were assigned from~\citep{james2013introduction,tan2006idm,rajaraman2011mining,murrell2010introduction,stanton2012ids}.

Concepts from the reading were developed further during the lecture periods in conjuction with implementations demonstrated in \R. Homework, consisting of conceptual questions requiring written responses as well as computational questions requiring coding in \R, was due approximately every two weeks. Two exams were given---both of which had in-class and take-home components. The first exam was given after the first two modules, and focused on data visualization and data manipulation principles demonstrated in written form. The second exam unfolded over two weeks, and focused on the difficult breast cancer classification problem discussed above. An open-ended project (see below) brought the semester to a close. More details on these assignments, including sample questions, are presented in our supplementary materials. 



\paragraph{Project}

The culmination of the course was an open-ended term project that students completed in groups of three. Only three conditions were given:

\begin{enumerate}
  \itemsep0in
  \item Your project must be centered around data
  \item Your project must tell us something
  \item To get an A, you must show something beyond what we've done in class
\end{enumerate}

Just like in other statistics courses, the project was segmented so that each group submitted a proposal that had to be approved before the group proceeded~\citep{halvorsen2001motivating}. The final deliverable was a 10-minute in-class presentation as well as a written ``blog post" crafted in \RM~\citep{markdown}. 

Examples of successful projects are presented in the supplementary materials.

\section{Feedback}

The feedback that I have received on this course---through informal and formal evaluations---has been nearly universally positive. In particular, the 42 students (mostly from \Smith but also including five students from three nearby colleges) seemed convinced that they learned ``useful things." More specific feedback is available in our supplementary materials. 

Some of these students were able to channel these useful skills into their careers almost immediately. Internships and job offers followed in the spring for a handful of students in the first offering: two students spent their summers at NIST, and one later accepted a full-time job offer from MIT's Lincoln Laboratory. External validation also came during the Five College DataFest, the inaugural local version of the ASA-sponsored data analysis competition~\citep{gould2014datafest}. DataFest is an open-ended data analysis competition that challenges students working in teams of up to five to develop insights from a difficult data set. A team of five students from \Smith---four of whom had taken this course---won the Best In Show prize. In this case, skills developed in the course helped these students perform data manipulation tasks with considerably less difficulty than other groups. For example, each observation in this particular data set included a \emph{date} field, but the values were encoded as strings of text. Most groups struggled to work sensibly with these data, as familiar workflows were infeasible (e.g. the data was too large to open in Excel, so ``Format Cells..." was not a viable solution). The winning group was able to quickly tokenize this string in \R, and---having jumped this hurdle---had more time to spend on their analysis.

\section{Discussion}

It is clear that the popularity of \emph{data science} has brought both opportunities and challenges to the statistics profession. While statisticians are openly grappling with questions about the relationship of our field to data science~\citep{davidian2013ds, davidian2013bd,franck2013is,bartlett2013we}, there appears to be less conflict among computer scientists, who (rightly or wrongly)
distinguish data science from statistics on the basis of the heterogeneity and lack of structure of the data with which data scientists, as opposed to statisticians, work~\citep{dhar2013}. As \emph{Big Data} (which is clearly related to---but too often elided with---data science) is often associated with computer science, computer scientists tend to have an inclusive attitude towards data science. 

A popular joke is that, ``a data scientist is a statistician who lives in San Francisco," but Hadley ~\cite{wickhamtweet}, a Ph.D. statistician, floated a more cynical take on Twitter: ``a data scientist is a statistician who is useful." 
Statisticians are the guardians of statistical inference, and it is our responsibility to educate practitioners about using models appropriately, and the hazards of ignoring model assumptions when making inferences. But many model assumptions are only truly met under idealized conditions, and thus, as~\cite{box1979some} eloquently argued, one must think carefully about when statistical inferences are valid. 
When they are not, statisticians are caught in the awkward position, as Wickham suggests, of always saying ``no." This position can be dissatisfying. 

If data science represents the new reality for data analysis, then there is a real risk to the field of statistics if we fail to embrace it. The damage could come on two fronts: first, we lose data science and all of the students who are interested in it to computer science; and second, the world will become populated by data analysts who don't fully understand or appreciate the importance of statistics. While the former blow would be damaging, the latter could be catastrophic---and not just for our profession. Conversely, while the potential that data science is a fad certainly exists, it seems less likely each day. It is hard to imagine waking up to a future in which decision-makers are not interested in what data (however it may have been collected and however it may be structured) can offer them. 


Data science courses like this one provide a mechanism to develop students' abilities to work with modern data, and these skills are quickly transitioning from desirable to necessary. 

\bibliographystyle{asa}
\bibliography{references}

\newpage
\appendix
\section{Additional Topics}
\label{sec:topics_app}

\paragraph{Spatial Analysis}

At \Smith, we have a Spatial Analysis Lab (SAL) staffed by a permament director and a post-baccalaureate fellow. These two people were particularly interested in the data science course, and attended most of the class meetings in the fall of 2013. We were able to build a partnership that brought the fascinating world of spatial analysis and GIS to \ds. Students from the class were encouraged to attend a series of workshops and lectures sponsored by the SAL over the course of the semester. 

The groundwork for the importance and interest in data maps was laid in the visualization unit through discussion of John Snow's cholera map, and other famous data maps presented by~\cite{tufte1983visual} and ~\cite{yau2013dp}. As the data manipulation unit winded to a close, an entire class period was devoted to spatial analysis and mapping. The SAL fellow delivered a lecture on the basics of data maps, providing examples and identifying key concepts in how data maps are created (e.g. normalization, projection systems, color scales, etc.). Next, we turned to how to work with spatial data: what are shapefiles? Where can they be found on the Internet? How can they be used in \R? How do you draw a choropleth in \R? Finally, students were given some data from the admissions office about the hometowns of \Smith students, and asked to create a choropleth in \R. 

Being familiar with only the basics of spatial analysis myself, I was astounded by the depth of the topic. Students also responded very positively to the prospect of creating colorful data maps. One student became so interested in working with ArcGIS that she began frequenting the SAL. In future incarnations of the course, one could consider extending this portion of the course into a full module. 

\paragraph{Text Mining \& Regular Expressions}

Not all data is numerical, and since the purpose of this course is to provide students with tools to work with a variety of data, one or two class periods were devoted to working with text. There are many interesting questions that can be asked of data stored as text~\citep{mosteller1963inference}, and many places to find such data. We focused on the works of Shakespeare, which are conveniently available through Project Gutenberg~\citep{gutenberg}. A simple question is: how many times does (the character) Macbeth speak in (the play) \textit{Macbeth}? 

The motivating challenge is that computers are highly efficient at storing text, but not very good at understanding it, whereas humans are really good at understanding text, but not very good at storing it. Thus, to answer our question, a human would have to scan the entire book and keep track of how many times Macbeth spoke. Identifying when Macbeth speaks is easy, but scanning the book is labor-intensive. It is also not a scalable solution (imagine having to do this for \emph{every} character in the book---or for \emph{all} of Shakespeare's plays instead of just one). Conversely, a computer can scan the entire book in milliseconds, but needs to be instructed as to the pattern that indicates that Macbeth is speaking. Pattern-matching through \emph{regular expressions}, available in the UNIX shell or in \R via the function \cmd{grep()}, provides an incredibly powerful mechanism for solving problems of this nature. 

While developing expertise with regular expressions is not likely to occur in just a few class periods, a basic understanding is within the grasp of most students. This basic understanding is enough to give the student \emph{confidence} that she could solve this type of problem in the future, if given enough time and resources. 

\paragraph{Data Expo}

Probably the most universally well-received aspect of the course was the Data Expo, which was held midway through the semester. At this point in the course, students were beginning to think about what they wanted to do for their term projects. With the idea of jump-starting their thinking, five members of the community (faculty from Engineering, staff from SAL, Institutional Research, and Administrative Technology, and a local data scientist working for a non-profit) came to talk to the class about the data with which they work, the questions they would like to answer with it, and the hurdles they currently face. The excitement in the room was palpable---one student told me afterwards that it was ``awesome" and that ``everyone is just so pumped right now to get started on their projects." Another student gushed about seeing the real-world applications of data science. 

One of my goals for this course was to convince students that there are two major kinds of skills one must have in order to be a successful data scientist: technical skills to actually do the analyses; and communication skills in order to present one's findings to a presumably non-technical audience. There were several moments during the Data Expo when the guest speakers underscored those very qualities, and expressed exasperation that so few people seemed to have both. For the instructor, this kind of third-party validation was invaluable at creating ``buy-in" among students. Students, all but one of whom were juniors and seniors, seemed energized at the prospect that what they were learning in class might translate directly into employable skills. 

\paragraph{Network Science}

During the last few weeks of class, students were hard at work on their projects (most of which were very demanding), and had limited bandwidth for absorbing new content. Thus, the final few class periods were devoted to special topics that were interesting, accessible, and fun, and would extend the main goals of the course. 

One of my own research fields is network science, which has both theoretical and applied aspects. Network science theory can be seen as an extension of the mathematical subject of graph theory, but applied to real-world networks in an attempt to form valid mathematical models for the networks that we see in reality. For example, Google's PageRank algorithm fits squarely into existing notions of centrality in graphs developed by discrete mathematicians and computer scientists, but it's application to the web and other real-world networks provokes interesting questions about the nature of communication networks in general. 

Students in \ds were exposed to these ideas and a few others. For example, students were asked to construct a ``Kevin Bacon Oracle" that would find the shortest string of movies connecting two actors or actresses using data from the Internet Movie Database (IMDB~\citep{imdb}). This requires understanding the network structure of Hollywood (i.e. actors/actresses are nodes, edges between nodes exist when both actors have appeared in a movie together), the notion of shortest paths in graphs, the structure of the IMDB, including how to write SQL queries to retrieve the relevant information, and how to use a graph library in \R (i.e. \cmd{igraph}) to pull all of it together. Projects of this nature are emblematic of the diverse sources of knowledge that make data science interesting. 

\paragraph{Big Data}

\ds is not a course on Big Data. Rather, it should be thought of as a precursor to Big Data~\citep{horton2015setting}, in that students leave data science with the experience of working with large data sets and knowledge of practical tools for doing so. The term ``Big Data" is not so well-defined---having both absolute and relative definitions that are popular. One relative definition that is relevant to most people---especially budding data scientists---is that ``Big Data is when your workflow breaks."\footnote{This definition of ``big data" has been popularized by Randall Pruim.} Students in \ds now understand why and when this type of difficulty arises, and have been given a series of tools to obviate that hurdle. However, they have not been exposed to Big Data in an absolute sense. On the last day of class, we discussed concepts that are most likely associated with absolute Big Data: parallel processing, non-SQL data storage schemes, Hadoop, and MapReduce. We reviewed the canonical MapReduce example of tabulating word frequency in a large number of documents, and ran Python and \R code to do the computation. 

Certainly, this is not enough time to go in-depth with Big Data. But after having absorbed so much over the past three months, and having had a chance to showcase their new skills through their projects, students leave with a sense of satisfaction about what they have learned, but with their appetites whetted for the frontier that lies ahead. 

\section{Projects}

These two abstracts summarize two of the more successful student projects in \ds:

\begin{itemize}
  \item weathR\footnote{This project received an Honorable Mention in the \href{https://www.causeweb.org/usproc/2014_WinningPrj.htm}{2014 Undergraduate Statistics Class Project Competition} (USCLAP).}: Twitter is an information product generated by its users. As a collection of records direct from the horse's mouth (or fingers, as it were), Tweets have a relationship with goings-on in users' lives. We can use Tweets to examine the relationship between what users post---their perceptions of the world around them---and conditions in the world. For example, data scientists have been investigating whether or not we can predict governmental elections and stock market changes using Twitter. We examined this connection between social media posts and the external world through a facet of users' lives that is accessible to us as data miners, universal and impactful: weather.
  \item \Smathies~\footnote{Math majors from \Smith refer to themselves as \Smathies.}: For undergraduates, the prospect of applying to graduate programs can be a daunting process, made more difficult if that student lacks connections to the institutions in which they are interested. The knowledge of an existing connection currently or previously in a program - whether as an advisor or an alumni - can provide an undergraduate with a valuable stepping-stone towards making a final decision about said program.

This project aims to explore the interconnectedness of \Smith College alumnae, specifically those who went on to earn a PhD in Mathematics or a related field. Our motivation for this project is to not only explore the connections that \Smith alumnae make, but to also develop a valuable tool for undergraduates searching for institutions or advisors with or without prior connections to \Smith. Specifically, the project explores connections between \Smathies and their advisors, \Smathies and their advisees, and \Smathies and their co-authors with whom they have written academic papers. The connections allow us to see how far the \Smathies world extends beyond \Smathies themselves. We can explore whether or not \Smathies have collaborated with a select group of co-authors, or if they prefer to seek out a variety of individuals. Additionally, we can see whether or not \Smathies had preferred advisors.

The point of this project, ultimately, is to explore both how connected \Smathies are with the rest of the Mathematics world and retrieve relevant information for current undergraduates.

\end{itemize}

\begin{figure}
  \centering
  \begin{subfigure}{0.45\textwidth}
    \includegraphics[width=\textwidth]{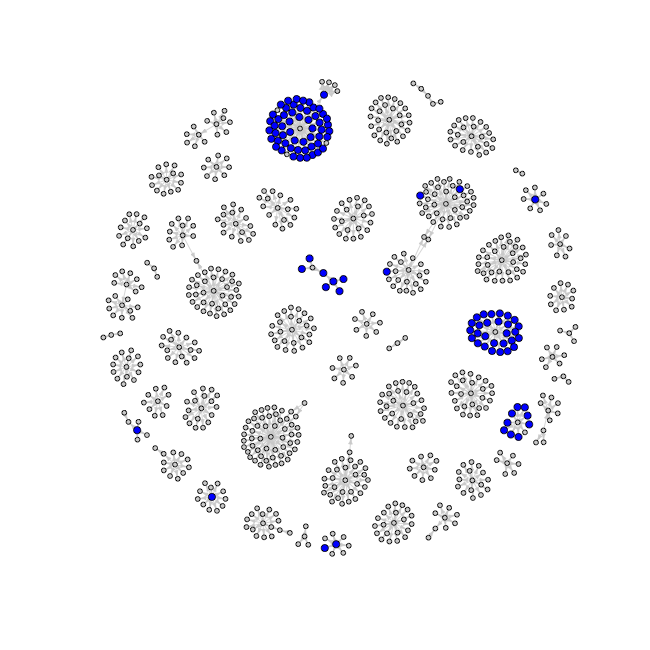}
  \end{subfigure}
  \begin{subfigure}{0.45\textwidth}
    \includegraphics[width=\textwidth]{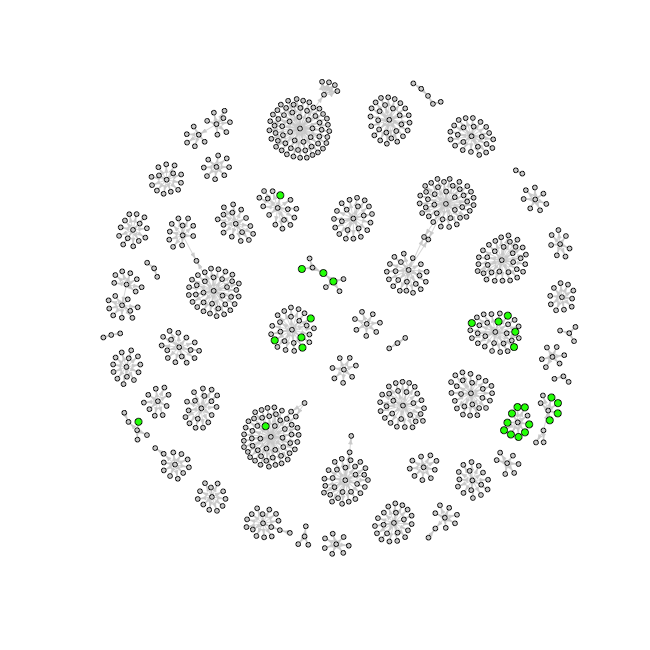}
  \end{subfigure}
  \caption{Visual depictions of the \Smathies alumni network creating by \ds students. Each node is an alum, or her dissertation advisor, and directed edges connect the two. On the left, color corresponds to the student's graduate school; on the right, their field. These are static images taken from a web application created with Shiny~\citep{shiny}. These figures might help one decide where to go to graduate school, by illuminating the pathways that previous graduates from \Smith have trod.}
  \label{fig:smath}
\end{figure}

Two visual depictions of the \Smathies network are shown in Figure \ref{fig:smath}.

\section{Feedback}

The following block quotations are excerpts from a Mid-Semester Assessment (MSA) conducted by the Jacobson Center for Writing, Teaching and Learning at \Smith College. An MSA Specialist meets with students without the instructor being present, and solicits responses to specific questions. The text of the report is written by the MSA Specialist, but phrases in quotation marks are quotations from actual students. Student responses are aggregated using a ``all/most/many/some/one" scale.

\begin{quotation}
\noindent All students agree that they are learning useful things:
``We are learning useful things--enough variety that even the most seasoned of us is still seeing new things...the examples are cool...we're learning a new, better way to think and ask questions...our projects teach us a way to tackle a problem."
\end{quotation}

To that end, the Data Expo was successful in helping students to appreciate the challenges people face when working with data. 

\begin{quotation}
\noindent All students enjoy ``Data Expo Day...the coolest!" and praise ``the guest speakers for us to learn more about data science in the real world"; one group adds, ``Thanks for the job opportunities and letting us see what one can do in jobs and life with this."
\end{quotation}

\section{A Note to Prospective Instructors}


Several people familiar with this course have asked about the skills required to teach this course. From my point-of-view the most important thing is to have the same willingness to learn new things that you ask of your students. In terms of the content, a deep knowledge of all subjects is not required, although comfort and troubleshooting ability with \R is necessary. Students are willing to accept a certain amount of frustration that goes hand-in-hand with learning a new programming language, but when they encounter roadblocks that seem immovable, that frustration can mutate into helplessness. The instructor must provide enough support mechanisms to avoid this. Student teaching assistants and office hours are especially helpful. 

Even without prior knowledge, enough of the material on data visualization and machine learning can be absorbed in a relatively short period of time by reading a few of the books cited. SQL has many subtleties that are not likely to come up in this course, but the basics are not difficult to learn, even via online tutorials and self-study. Here again, some experience and practice are important. 

For students, prior programming experience is essential. Experience with \R is not required, and in my experience, computer science majors with weaker statistical backgrounds usually fared better than students with stronger statistical backgrounds but less programming experience. This was a demanding course that required most students to spend a substantial amount of time working through assignments. However, even students who struggled were so convinced that what they were learning was useful that there were few serious complaints. Nevertheless one could certainly experiment with slowing down the pace of the course. 

\newpage
\section{Sample Exam Questions}

\begin{enumerate}

  \item (20 pts) Suppose that we have two rectangular arrays of data, labeled $students$ and $houses$. [In SQL terminology, $students$ and $houses$ are both \cmd{table}s. In \R terminology, they are \cmd{data.frame}s.] $students$ contains information about individual XXXXX students (e.g. her student id, name, date of birth, class year, campus house, etc.). Each row in $students$ contains data about one student. $houses$ contains data about XXXXX houses (e.g. house name, capacity, street address, etc.). Each row in $houses$ contains data about one XXXXX house. 
  
Suppose further that we want to generate a student address book. The address book will consist of two columns of data: the first column will contain the student's name; and the second will be the address where she lives. 	
	
	\begin{enumerate}
		\itemsep2in
		\item Describe, \textbf{in words}, a data management operation that you could perform in order to achieve this. Be as specific as you can about what the operation will do and how it must be specified, but note that you do \textbf{not} have to write or reference SQL or \R code!

		\item It is important that every student appears in the address book, regardless of whether she lives on campus. Would a JOIN, LEFT JOIN, or RIGHT JOIN be most appropriate? Explain why.

	\item Suppose now that only students from the Class of 2014 are to be included in the address book. What additional data management operation could you perform to achieve this? Again, be specific, but there is no need to write code. 
	
	\end{enumerate}

  \newpage
  
	\item (10 pts) Briefly discuss the relative strengths of SQL vs. \R. What does SQL do better than \R? What does \R do better than SQL? [Hint: It may be helpful to give an example of a data science task for which one or the other would be better suited.]

\newpage
  
  \item (20 pts) You are working for the National Transportation Safety Board as a traffic engineer. One of your colleagues has written an \R function that will simulate virtual traffic through the toll booth at the Holyoke/Springfield exit on the Massachusetts Turnpike. Given initial conditions based on the day of week and the temperature, the simulator will randomly generate traffic, and record waiting times (in minutes) for individual cars at this toll booth. Thus, you might use the function like this:

\begin{knitrout}
\definecolor{shadecolor}{rgb}{0.969, 0.969, 0.969}\color{fgcolor}\begin{kframe}
\begin{alltt}
\hlkwd{simulate.traffic}\hlstd{(}\hlkwc{dayofweek} \hlstd{=} \hlstr{"Thursday"}\hlstd{,} \hlkwc{temp} \hlstd{=} \hlnum{48}\hlstd{)}
\end{alltt}
\begin{verbatim}
##  [1] 1 2 1 0 1 1 0 0 0 1 1 1 2 1 0 0 0 2 1 2 1 1 1 0 0
\end{verbatim}
\end{kframe}
\end{knitrout}

  \begin{enumerate}
    \itemsep2in
    \item Describe how you could use this simulator to find a confidence interval for the mean waiting time for Mondays with a temperature of 56 degrees.
    \item Describe how you could use this simulator to find a confidence interval for the mean waiting time for an average day (i.e. a generic day, regardless of which day of the week or what the temperature was).
  
  \end{enumerate}
  
\end{enumerate}

\end{document}